# Accurate spectra for high energy ions by advanced time-of-flight diamond-detector schemes in experiments with high energy and intensity lasers

M. Salvadori[1,2,3]*, F. Consoli[3], C. Verona[4], M. Cipriani[3], M. P. Anania[5], P. L. Andreoli[3], P. Antici[2], F. Bisesto[5], G. Costa[1,5], G. Cristofari[3], R. De Angelis[6], G. Di Giorgio[3], M. Ferrario[5], M. Galletti[7], D. Giulietti[6,8], M. Migliorati[1,9], R. Pompili[5], and A. Zigler[10]

[1]Università d Roma La Sapienza, Piazzale Aldo Moro 5, Roma, Italy
[2]INRS-EMT, Varennes, Québec, Canada
[3]ENEA Fusion and Technologies for Nuclear Safety Department, C.R. Frascati, Via Enrico Fermi 45, Frascati (Roma), Italy
[4]University of Rome "Tor Vergata", Industrial Engineering Department, Roma, Italy
[5]INFN-LNF, Frascati, Roma, Italy
[6]Department of Physics, University of Pisa, Largo Bruno Pontecorvo 3, Pisa, Italy
[7]GoLP Instituto de Plasma e Fusão Nuclear, Istituto Superior Técnico, Universidade de Lisboa, Lisbon, Portugal
[8]INFN of Pisa, Largo Bruno Pontecorvo 3, Pisa, Italy
[9]INFN sezione di Roma1, Piazzale Aldo Moro 2, Roma, Italy
[10]Racah Institute of Physics, Hebrew University, Jerusalem, Israel

*Corresponding author:
**Martina Salvadori**
Email: martina.salvadori@enea.it

ABSTRACT: Time-Of-Flight (TOF) methods are very effective to detect particles accelerated in laser-plasma interactions, but they shows significant limitations when used in experiments with high energy and intensity lasers, where both high-energy ions and remarkable levels of ElectroMagnetic Pulses (EMPs) in the radiofrequency-microwave range are generated. Here we describe a novel advanced diagnostic method for the characterization of protons accelerated by intense matter interactions with high-energy and high-intensity ultra-short laser pulses up to the femtosecond and even future attosecond range. The method employs a stacked diamond detector structure and the TOF technique, featuring high sensitivity, high resolution, high radiation hardness and high signal-to-noise ratio in environments heavily affected by remarkable EMP fields. A detailed study on the use, the optimization and the properties of a single module of the stack is here also described for an experiment where a fast diamond detector is employed in an highly EMP-polluted environment. Accurate calibrated spectra of accelerated protons are presented from an experiment with the femtosecond Flame laser (beyond 100 TW power and ~$10^{19}$ W/cm$^2$ intensity) interacting with thin foil targets. The results that can be readily applied to the case of complex stack configurations and to more general experimental conditions.

Thanks to the development of new high energy and high-power lasers[1], during the last decades the field of laser-plasma particle acceleration has experienced a significant evolution and expansion[2-4]. The interest in this field arises from the possibility to produce, with a compact system, intense bunches of high energy charged particles that can be used for several notable applications[5,6], including inertial confinement fusion[7]. According to the laser parameters, different acceleration processes can take place during the laser-matter interaction[2-4]. In the Target Normal Sheath Acceleration (TNSA) a high-power laser (up to the PW power and to the $10^{21}$ Wcm$^{-2}$ intensity) is focused on a solid target, leading to hot electrons generation at the front side. These electrons can penetrate the target, reach its rear side and escape from it. A strong sheath field is created due to charge separation on the rear side of the target. Protons and heavier ions are thus accelerated by the remarkable electric fields and follow the electron expansion in the vacuum chamber[2-4].

The study of the acceleration processes that take place in the frame of the laser-plasma interaction requires the development of dedicated diagnostics able to characterize the properties of the accelerated particle beams and thus to outline the differences among the different possible acceleration mechanisms. These diagnostics have to meet some important requirements, depending on the characteristics of the accelerated particle beams. Among these, high sensitivity and high energy resolution are needed for the accurate spectrum reconstruction of the accelerated ions. Moreover, the notable dependence of these beams on the angle of emission requires a precise characterization of the particle angular distribution at different directions. This can be done by methods employing stacks of passive detectors as for example Radiochromic films (RCF)[8,9], which allow to reconstruct the beam divergence, but have an intrinsic low energy resolution and do not distinguish among different ion species. Alternatively, electrostatic-magnetostatic spectrometers of the Thomson type[9,10] are able to recognize the different charge-over-mass ratios for the incoming ion species and to reconstruct their spectra, but do not have any angular discrimination. The physical dimension of these devices usually gives notable difficulties to place several of them at different directions, on the purpose to retrieve information on the angular beam emission, especially in the cases where the vacuum chamber is of limited dimension. An additional requirement that the ideal diagnostic system has to satisfy is the capability to work at high repetition rates (i.e. up to kHz), so without having to open the experimental chamber after each shot. This task cannot be fulfilled when passive detectors such as RCF, CR39 or Imaging Plate are used, both in the stack configuration or within Thomson spectrometers.

Time-of-Flight (TOF) technique is commonly adopted in laser-plasma experiments to obtain time-resolved characterization of the accelerated ions[9,11-14]. This method has the potential to meet all the discussed requirements. The reduced dimension and compactness of some TOF detectors allows to install several of them along different lines of observation to get suitable information on the angular beam emission. The possibility to work with real-time readout systems, without the need of opening the chamber after each shot, makes the TOF an ideal candidate for ion diagnostics in high repetition-rate experiments. Energetic ions are usually detected using semiconductor devices made of SiC or diamond[15-17], or scintillators[18,19].

Nevertheless, the method has historically shown significant limitations when applied to experiments with high energy and intensity lasers, where both high-energy ions[20,21] and remarkable levels of Electromagnetic pulses (EMPs) in the radiofrequency-microwave range[22] are generated. Protons up to several tens of MeV are here produced[23], with typical spectra decreasing with energy. So, both high sensitivity and high resolution are required for a suitable description of the most energetic part of the ion emission. Chemical Vapor Deposition (CVD) diamond detectors have the potential to satisfy both these needs, but with the TOF technique energy resolution and sensitivity have somehow opposite requirements. Indeed, improved energy resolution is here commonly obtained in two ways.

1) By using fast detectors. In TOF, energy resolution is linked to time resolution, and the detector response velocity is mainly limited by the thickness of the active sensitive region, associated with the time of charge collection to the electrodes. The typical monocrystalline diamond detector response to a single 5.486 MeV α particle emitted by $^{241}$Am decay can have FWHM up to 8 ns for a detector with 500 µm thickness[24], up to 1.1 ns for one with just 100 µm thickness[24] and up to 0.8 ns or below for one with 50 µm thickness[17,25]. So thinner detectors should be preferred for better energy resolution, but on the other hand energetic protons will pass through them more easily, leaving there only a small amount of their energy, and this will lead to reduced sensitivity of the detector to them. In particular, it was shown by Monte-Carlo SRIM code simulations[26] that protons with energies up to $E_{THRS-100}$ ~ 4.6 MeV will be stopped within the active region of the 100 µm diamond, releasing all their energy in the detector. Alternatively, those with much higher energy will lose a very limited portion of it. If the signal produced in the detector is small with respect to the background noise, the information on the associated particles can be lost. As observed, the emitted proton spectrum in these experiments decreases with energy, and thus this problem is even more significant for the most-energetic protons. For particles passing through the detector whose produced signal is still reasonably higher than the background, it is possible to reconstruct, by means of suitable numerical simulations[13], the initial particle energy from the small portion released in the detector and thus estimate the associated number of detected particles. But this will commonly suffer of low precision due to the related large confidence interval. So, high energy resolution will be achieved but with low resolution for the number of particles in the spectrum.

2) By placing the detectors at larger distances from the target. However, this reduces the overall sensitivity by decreasing the solid angle of detection and thus the number of particles intercepted by the detector.

For these reasons, whenever improved sensitivity for the most energetic ions is required, thicker detectors are used[13,24,27], but at expenses of the time resolution[24], and thus of the related energy resolution accordingly. One key issue for the sensitivity threshold is the background noise associated with the technique. This is due to the intrinsic noise of all the electronic equipment used, and in particular of the oscilloscope, and to the external electromagnetic noise coupled to the overall TOF detecting chain. It is well known that in energetic high power laser facilities the intense laser-matter interaction generates strong Electromagnetic Pulses (EMPs)[28-33] in the radiofrequency-microwave range, which remarkably affect the signal-to-noise ratio of the electronic devices placed inside and nearby the vacuum chamber, and that commonly scales with the laser intensities[33,34]. Since TOF measurements require time-resolved detectors connected with fast electronic readout systems, the presence of remarkable levels of EMPs has, so far, strongly hindered the employment of this technique in experiments with intense ultra-short pulses at very high-power regimes, where instead their features would have been very useful.

In this work we present a novel advanced diagnostic method for ions generated by the interaction with matter of high energy and high intensity laser pulses with a time duration in the femtosecond or attosecond range, based on time-of-flight technique and ensuring high sensitivity, high resolution and high signal-to-noise ratio in environments heavily affected by remarkable EMP fields. The method aims at retrieving accurate calibrated spectra of laser-plasma accelerated protons, developing and exploiting the capabilities of the TOF measurements in combination with those of the CVD diamond detectors.

## METHODS

**Detector principle**
The basic idea is to use a composite structure made of a set of several detectors, which can be positioned as a stack, one after to the other along the direction of the impinging ions emitted from the laser-matter interaction, as shown in Figure 1a. Filters of different materials and thicknesses can be also used in front of the detectors to condition the incoming particle beams. Here we consider diamond detectors only, but the methodology can be applied also to solid-state detectors made of other semiconductors, such as for instance SiC detectors. The thickness of each detector in the stack can be chosen to optimize the performances of the whole sensor. The total length would be given by the sum of all the detector thicknesses in the stack. In particular, the use of thin (i.e. 50-100 µm) modules could give high energy resolution and nice sensitivity to high-energy protons, at expenses of a larger number of modules in the stack. In addition, using thin diamond layers allows for a high radiation hardness of the whole diamond detector[35,36].

The main advantage of this novel design is the capability to detect high-energy particles with good sensitivity, without sacrificing energy resolution. The sensitivity features of thick diamonds can be thus replaced with the layered multiple thin diamonds, insuring fast time response and high radiation hardness. It is worth to underline that the technology of diamond detectors is already mature for such complex structures. Devices based on multiple diamond detectors are reported in the literature for other applications: proton recoil telescope based on two diamonds for fast neutrons diagnostic[37] and monolithic diamond-based ΔE-E charged particle telescope for identification of fragments coming from the nuclear reactions[38].

Whenever these schemes will be applied to detect ions emitted by intense matter interactions, with high energy and intensity laser pulses up to the femtosecond and attosecond time-scales, the main limitations to the method sensitivity will be given by the coupling of giant EMPs (up to several MV/m levels[22]) to the related electronic devices. This is known to be a major issue especially for devices working at larger frequency bandwidths[22] and will thus be more serious for the fast (i.e. thin) detectors. The capability to deal with this fundamental issue is one of the key factors for the successful and effective implementation of these methodologies in the future and will be thus here described in detail.

This work is meant as the advanced developing of a methodology capable to fully take advantage of the properties of the TOF methods applied to diamond detectors. For this reason we describe here a detailed study on the use, the optimization and the properties of a single module of the stack, in experiments with high energy and high intensity laser pulses with femtosecond time-duration interacting with thin foil targets, on the purpose to apply the associated considerations to the case of complex stack configurations and to more general experimental conditions.

**Description of the single-module diamond**
As mentioned, we consider that the single module of the stack is a CVD monocrystalline diamond detector. It was developed and optimized to work in the harsh environments of high-power and high-energy laser-matter interactions[17], where high levels of EMP fields are generated. Thanks to their wideband gap of 5.5 eV, their high carrier mobility and their radiation hardness[35,36,39,40], they are particularly useful to monitor UV, X photon emission and accelerated particles in laser-plasma experiments[17,36,41-43]. Indeed, they have low leakage current, fast time-response, and insensitivity to visible and infrared light which avoids any effect due to possible coupling of any stray light coming from the main laser beam to the detector. This is important to get an intrinsic low background level. To fully investigate the management of the EMP issue in the most demanding conditions, we considered here a fast detector consisting of a 50 μm intrinsic diamond layer grown by chemical vapour deposition technique on a commercial 4 × 4 × 0.5 mm High Pressure High Temperature (HPHT) substrate. The detector, shown in Figure 1b, operates in planar configuration with superficial interdigital aluminium contacts with 20 µm width and 20 µm spacing[17], which allow for fast time detection of moderate energy particles. The time resolution of this detector was offline obtained from detection of single α particles of $E_\alpha$ = 5.486 MeV energy, coming from $^{241}$Am source, and it resulted $\Delta t$ = 0.8 ns[25]. By considering a Gaussian fitting of the associated single-particle detected signal, this means a 3-dB frequency bandwidth $f_{3dB-diamond}$ = 625 MHz.

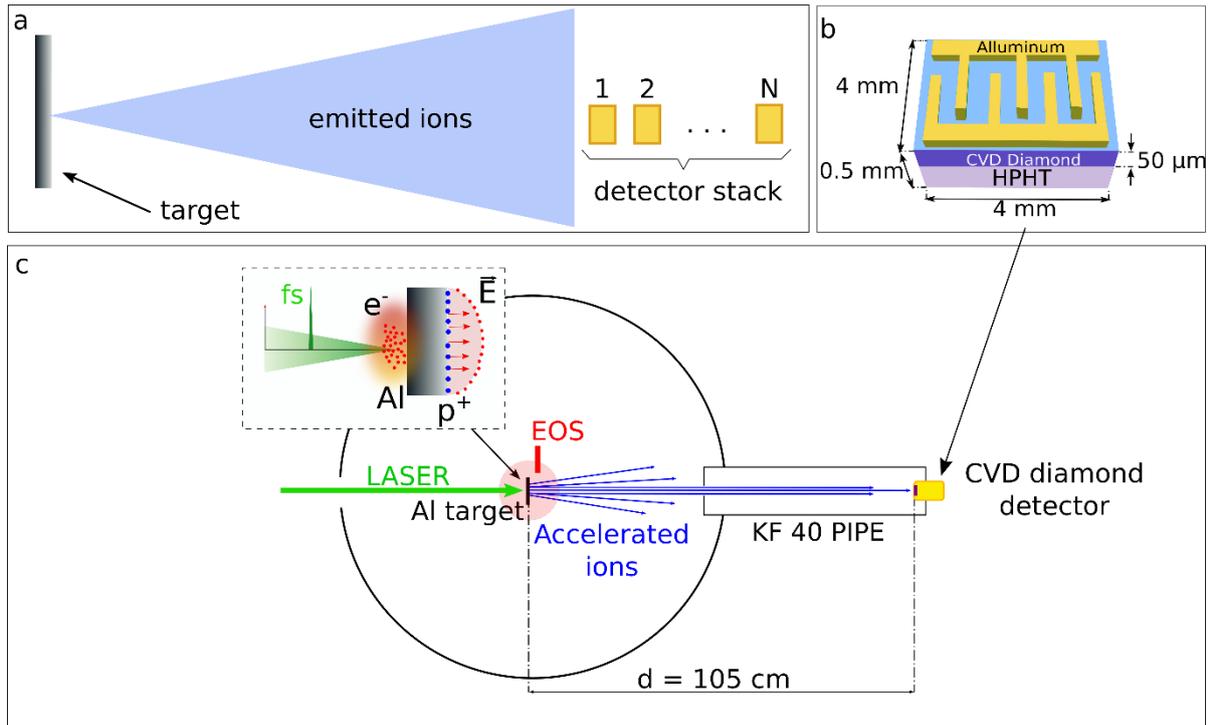

**Figure 1.** (**a**) Basic scheme of the advanced TOF detector. (**b**) Schematic representation of the diamond detectors in planar-interdigital configuration. (**c**) Experimental chamber layout and in the inset: a scheme of the laser-plasma interaction highlighting the TNSA mechanism, E is the electric field generated on the rear side of the target by the escaping hot electrons.

**Experimental Setup**

The diagnostic methodology was tested and optimized during an experimental campaign performed on 10 μm Aluminium flat targets, irradiated with the FLAME Ti:Sapphire laser at INFN, Laboratori Nazionali di Frascati, having 800 nm fundamental wavelength, 3 J maximum energy, 30 fs pulse duration and 25 µm focal spot, for 100 TW peak power and $2 \times 10^{19}$ Wcm$^{-2}$ maximum intensity on target[44]. The targets were fixed on an ad-hoc multi-target holder. During the experiment two main diagnostics were employed. An Electro-Optical Sampling (EOS) device was used to measure the longitudinal profile of the accelerated electron bunches[9,45-47] Accelerated ions were detected by a CVD diamond detector mounted at the end of a 105 cm long TOF line placed behind the target, as shown in Figure 1c.

**Calibrated spectrum determination: from data acquisition to spectrum reconstruction**

The determination of accurate energy spectra for incoming particles from the TOF signals generated by the detector is a delicate issue. In some cases, simple mathematical reasoning is used to get the energy-domain spectra from the stored time-domain signals, as very recently reported[13]. Anyway, they do not consider the specific experimental response of the detector to a specific particle with a specific energy.

We describe here instead a detailed procedure for getting accurate spectra from a diamond detector, including the experimental calibrations of its response. First of all, a detailed description of the acquisition system used to have a high dynamic range measurement with a nearly absent EMP noise is reported. Then we show the procedure used to include in the results the effects due to the long cables employed to manage the EMP pollution affecting the diamond signal. Finally, a novel methodology is described to get the calibrated energy spectrum. To this purpose, we analyze the response of the diamond detector and discuss in detail also the overall spectrum accuracy for the energy and the number of particles.

**High dynamic-range detection**

The TOF measurements employing diamond detectors are intrinsically characterized by large dynamic-range signals. In order to retrieve all the useful information from them, to accurately estimate the energy of the incoming particles and to finally compute their spectrum, it is necessary to distinguish the finest details without cutting the most intense portion of the signal. This is necessary in order to fully exploit the high sensitivity of the detector. A typical TOF measurement obtained in these cases is shown in Figure 2b. It mainly consists of two parts: the *photopeak* and the *particle contribution*.

The first is generated by the ionizing electromagnetic radiation emitted during the laser-matter interaction and, depending on the specific experimental conditions, can be higher or lower than the second. The photopeak is a clear and very useful signature of the laser-matter interaction instant. Its temporal position gives a reliable absolute reference for determining the arrival times-of-flight of particles from the target to the detector, as discussed in detail later.

The second main peak comes instead from the time superimposition of particles arriving to the detector. Both ions and electrons can produce a signal, but the electrons arriving at the same time instants of the ions have much lower energies, and thus their contribution result typically negligible with respect to them. Anyway, whenever required, the application of a very low magnetic field close to the diamond is usually sufficient to deflect them and to avoid their arrival to the detector, practically without modifying the trajectories of the ions.

In order to obtain complementary information from the two parts of the detected spectrum, the signal collected from the diamond detector was divided in two parts by a calibrated splitter, both having the same shape but half of the original amplitude. They were acquired by two different channels of the LECROY HDO 4104 scope (1 GHz bandwidth, 10 Gs/s sample rate per channel, 12-bit resolution, 8.4 Effective Number Of Bits (ENOB)), set with different amplitude scales. The use of this high-resolution and high-sensitivity oscilloscope is important for the improvement of the dynamic-range of the acquisition. The acquisition set-up is schematically depicted in Figure 2a. The intrinsic dynamic range of this specific low noise oscilloscope is indeed already high, but with this technique it was possible to improve it of a factor ~1.4. This configuration allowed to obtain an actual dynamic range of ~74 dB. Of course, the overall dynamic range obtained with this technique can be higher when oscilloscopes with lower intrinsic noise are used.

The signal reported in Figure 2b was obtained for the shot #38, with 2.5 J energy and $1.7 \times 10^{19}$ Wcm$^{-2}$ intensity. It is evident that channel one (Ch1 in the Figure) provides information about the main signal generated inside the diamond but, because of the finite dynamic range of the scope, there is poor resolution on the smaller details which, instead, can be obtained from channel three (Ch3 in the figure) which was set to a finer scale. Since the two channels of the same oscilloscope are accurately synchronized, from Ch3 the temporal position of the photopeak was retrieved and used as absolute reference for data shown in Ch1. The consistency of this comparison is confirmed by the coincident temporal position of the protons maximum energy given by the initial point of the main signal (yellow dotted line in Figure 2b).

The diamond was linked to the splitter through a cable connected with a custom large-capacity, high-frequency Millimetrica Bias-Tee (MIL-BT 010-6000). This isolated the scope from the -150 V bias voltage for the diamond, without losing significant information over the frequency range (300 kHz – 6 GHz), suitable and tailored for signals coming from the diamond detector.

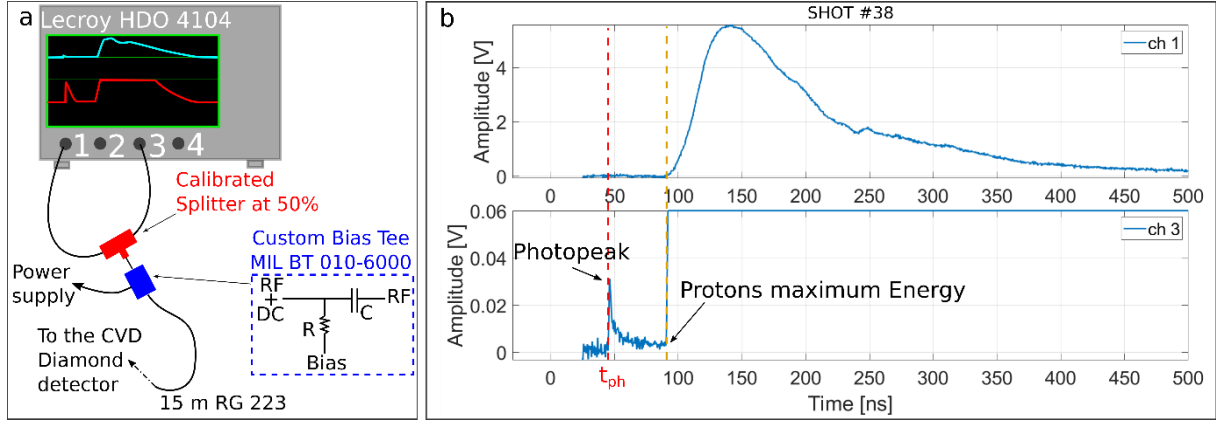

**Figure 2.** (**a**) The acquisition system configuration for the CVD diamond detectors and (**b**) a typical TOF signal acquired with the described system on channel 1 (Ch1) and channel 3 (Ch3) of the oscilloscope.

**EMP management and reduction**

It is well known that in high power laser-matter interactions the effect of the EMPs on diagnostics with electronical components can be remarkable[28-33]. Several precautions were thus here adopted to obtain signals from the diamond detectors with high signal-to-noise ratio even in these environments with high electromagnetic pulse pollution. In particular, two types of EMP field coupling had to be avoided: that directly with the detector and that with the overall acquisition system.

The EMP coupling with the device was minimized thanks to an ad-hoc mounting structure[17] of the diamond detector holder shown in Figure 3a. The diamond is mounted inside a compact cylindrical metallic enclosure with a minimal circular aperture on the front. The small radius $r$ provides a cut-off frequency, $f_{cutoff} \propto c/r$, where $c$ is the speed of light, which highly limits the EMP coupling without covering the active surface of the diamond detector[48]. The same principle was applied to the TOF line. The final connection with the diamond detector is a pipe of 65 cm length and $R_{TOF} = 20$ mm internal radius, as shown in the main picture of Figure 1. This acted as a cylindrical waveguide, and in particular for the first waveguide mode $TE_{11}$ the cut-off frequency was: $f_{cutoff} = c\,\dot{\chi}_{1,1}/(2\pi R_{TOF}) = 4.395$ GHz, where $\dot{\chi}_{1,1} = 1.841$ is the first root of the derivative of the first Bessel Function. The field intensity associated with a given mode decreases along the waveguide longitudinal direction $z$ with dependence $e^{-\alpha z}$. The attenuation factor provided by the pipe as a function of the frequency is shown in Figure 3b. A very good rejection to EMP fields traveling in the vacuum chamber was thus achieved.

In order to reduce the EMP coupling with the acquisition system, the main transmission link consisted of 15 m RG223 double-shielded coaxial cables, used to transport the signal from the detector to the scope. They have a shielding effectiveness higher than 80 dB for the frequency ranges related to the performed experiments and resulted very suitable to suppress the direct transmission of the EMP fields traveling in air to the inner coaxial conductor of the cable, through their outer shielding conductor. The coaxial cables together with the calibrated splitter and the custom bias-tee, were offline characterized in the frequency domain by the Agilent N5230A Vector Network analyser. The obtained $S_{21}$ scattering parameter is given in Figure 4a.

These long cables worked on the EMP rejection in several ways.

- They behave as effective low-pass filters[31], giving high attenuation to possible contribution at high frequencies. On one side this improved the filtering of the high-frequency components of possible EMPs coupled to the detector. But on the other side it might have introduced limitations to the fast response of the detector setup. As observed, the 3 dB low-pass bandwidth of the interdigital diamond detector resulted 625 MHz. According to Figure 4a, the $S_{21}$ scattering parameter for this frequency was 0.2492 (~12 dB), which indeed resulted a very suitable compromise.
- The propagation of the EMP fields outside the experimental chamber can be classically modelled with the 1/r law. The use of long cables allowed the positioning of the scopes far from the experimental chamber, and thus decreased the possible direct coupling of EMP fields with the scopes themselves.
- EMPs have a typical exponentially-decreasing time profile and become usually comparable to the background noise in ~100 ns. Since signals traveling in cables have lower velocity than those traveling in air, the cables were also used to introduce a temporal delay of several tens of nanoseconds between the high EMP contribution and the detected signal[31].

Those cables were also surrounded by several ferrite toroids, with bandwidths 0.5-5 MHz, 1 MHz-1 GHz, 2-30 MHz, 20-200 MHz, respectively, providing damping of the current flowing on the outer conductor of the cables caused by the applied external EMP fields. This avoided that those currents could reach the oscilloscopes and generate EMP noise on it.

By means of these methods it was possible to obtain a detected signal with high signal-to-noise ratio, without any numerical filtering procedure or the use of expensive high-quality Faraday cages, as clearly shown in Figure 2,4 and 6.

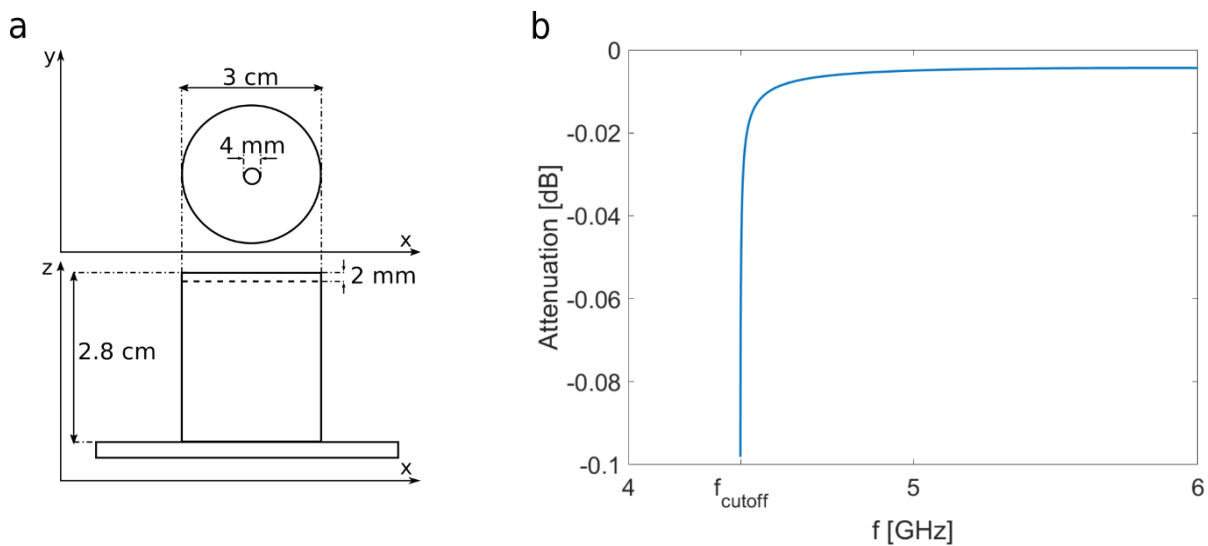

**Figure 3.** (**a**) Scheme of the EMP-optimized case of the diamond detector. (**b**) Attenuation of the $TE_{11}$ mode provided by the 65 cm pipe as a function of the frequency, with indication of the cutoff frequency.

**De-embedding procedure**

The signal collected by the CVD diamond detector was transmitted to the scope through the long coaxial link, having a frequency-dependent attenuation. Thus, a suitable de-embedding procedure[32] was required to recover the signal at the detector site from the one recorded on the oscilloscope. The de-embedded signal $S_D$ can be computed by the following relation[32]:

$$S_D(t) = \mathcal{F}^{-1}\left[\frac{\mathcal{F}(V(t))}{S_{21}(\omega)}\right] \quad (1)$$

where $V(t)$ is the signal stored on the scope and $\mathcal{F}$ and $\mathcal{F}^{-1}$ are the Fourier transform and the inverse Fourier transform operators, respectively. As a preliminary step, the Fourier transform of the signal and that of the background noise were computed and compared (see Figure 4b) to determine the bandwidth where a suitable signal-to-noise ratio was actually achieved. With the assumption of white-noise condition, the information about the background noise was retrieved from the portion of the signal collected in the time interval before the photopeak detection instant, when the laser-matter interaction is not occurred yet. Then, the frequency where the amplitude of the Fourier transform of the noise was equal to the Fourier transform of the signal, was selected as threshold frequency $f_c$ (highlighted in Figure 4b) and used to identify the frequency range in which the de-embedding procedure could be suitably applied. This preliminary step was necessary to avoid numerical amplification of the noise, and therefore equation (1) was applied only to the meaningful part of the acquired signal.

In Figure 4c the final de-embedded signal $S_D$ is obtained for the shot #38 and compared with the original raw signal V(t); it is evident that the contribution of the transmission line cannot be ignored.

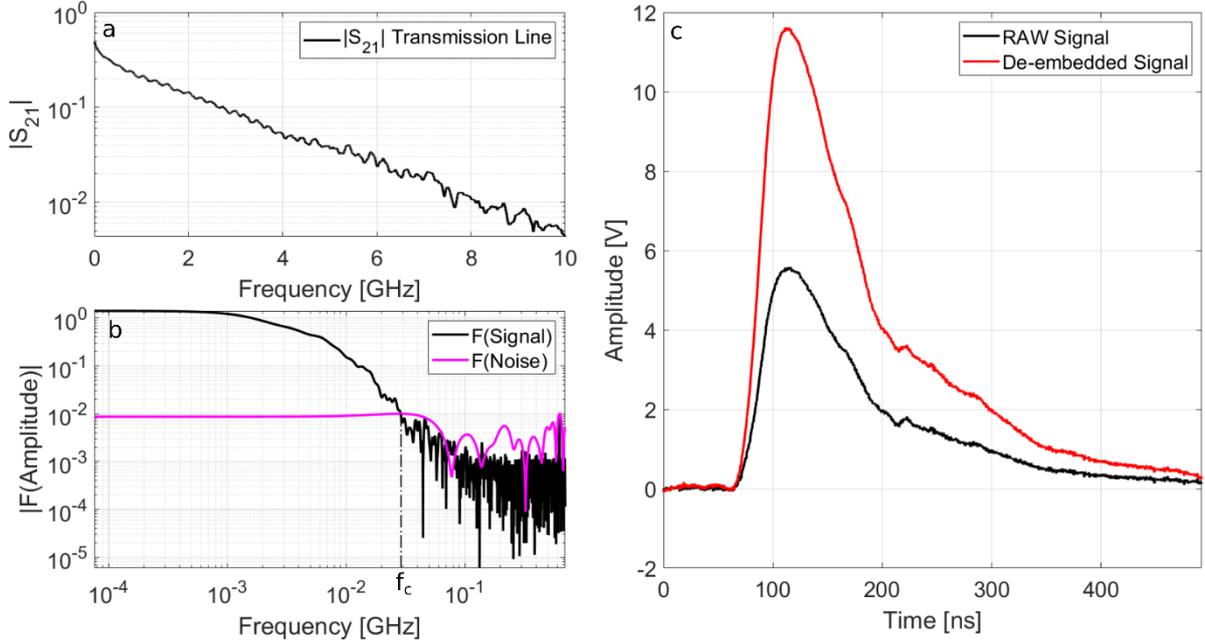

**Figure 4.** (**a**) The Scattering parameter $S_{21}$ of the transmission line (coaxial cables, splitter and bias tee) measured with the Agilent N5230A Network Analyzer (**b**) Fourier transform of the acquired signal and of the background noise (**c**) De-embedded signal $S_D$ compared with the original raw V(t), equal to the double of the signal detected on Ch3

**Analytical spectrum computation**
A. Particle energy estimation
To retrieve the accelerated particles spectrum, as a first step it is necessary to define an absolute time-reference related to the laser-matter interaction instant. As discussed, at the moment of laser interaction with matter, a burst of UV-X rays is generated and detected by the diamond detector. The detection time $t_{ph}$ of this narrow photopeak, shown in Figure 2b for shot #38, can be thus used as absolute time reference. The time needed for photons to travel from the source to the detector is $\Delta t_{prop} = d/c$, being $d$ the distance of the detector from the interaction point. Therefore, the time instant when the laser-matter interaction occurs can be determined as $t_{bang} = t_{ph} - \Delta t_{prop}$, and used as reliable absolute reference. We indicate with $t_i$ the ion detection time according to the new temporal reference and with $v_i = d/t_i$ the associated detected flight velocity. In this way, by TOF methods accurate information on particle velocity can be achieved. From these measurements calibrated particle spectra can be obtained, if the type of particle reaching the detector is known or assumed for a given time interval[25]. Under this condition, for a generic ion species the energy associated with the detection time can be obtained from the relation: $E_i = (\gamma_i - 1)m_i c^2$, where $\gamma_i$ and $m_i$ are the ion associated relativistic factor and its mass, respectively.

The finite temporal resolution $\Delta t$ of the detector, obtained with the mentioned calibration, affects the energy tolerance of the measurement performed with the diamond detectors. For a generic time interval, indicated with the $n$ index, with $t_n^i$ and $t_n^f$ initial and final extremes, will be $\Delta t = t_{i,n}^f - t_{i,n}^i$ and $E_{i,n}^i$ and $E_{i,n}^f$ the associated ion energies. Thus, the error on the energy estimation will be given by:

$$\Delta E_{i,n} = E_{i,n}^f - E_{i,n}^i = \frac{\partial E_i}{\partial \gamma} \Delta \gamma_i = m_i c^2 \frac{\partial \gamma_i}{\partial t} \Delta t \qquad (2)$$

The relative error on energy, given by the finite temporal resolution, can be written as:

$$\frac{\Delta E_{i,n}}{\bar{E}_{i,n}} = -\frac{m_i c^2}{\bar{E}_{i,n}} \left(\frac{c\Delta t}{d}\right) \left[\left(\frac{\bar{E}_{i,n}}{m_i c^2} + 1\right)^2 - 1\right]^{3/2} \qquad (3)$$

where $\bar{E}_{i,n} = \sqrt{E_{i,n}^f E_{i,n}^i}$ is the average ion energy in the $n^{th}$ interval. The negative sign is because energy is a monotonically decreasing function of time. In case of non-relativistic particles $\bar{E}_{i,n} \ll m_i c^2$, and the previous relation can be simplified obtaining:

$$\frac{\Delta E_{i,n}}{\bar{E}_{i,n}} = -2\sqrt{2} \frac{\Delta t}{d} \sqrt{\frac{\bar{E}_{i,n}}{m_i}} = -2 \frac{\Delta t}{\bar{t}_{i,n}} \qquad (4)$$

Where $\bar{t}_{i,n} = \sqrt{t_{i,n}^f t_{i,n}^i}$ is the average time in the $n^{th}$ interval. Thus, the absolute value of the relative error on the energy is equal to twice the relative error on the measurement of the time of ion arrival to the detector. So, for the same $\Delta t$ mainly due to the time resolution of the detector, the accuracy will be better for particles arriving to the detector at later times, so for detectors placed at larger distances from the target. But this, on the other hand, would decrease the number of particles hitting the same detector, and thus the signal amplitude.

B. Determination of the particle number.

As already mentioned, the interdigital diamond configuration consists a 50 µm intrinsic diamond layer, capable to fully stop protons with energy up to ~3 MeV. The signal generated by incoming protons is proportional to the particle energy. The induced charge signals collected at the sensitive electrode can be considered as due to the motion of electron/hole pairs generated in the bulk of the diamond sample, and charge pulses are formed only in regions where an electric field (E) is present. For this reason, the region where the charges are more efficiently collected is that closer to the electrode surface. As a consequence, the response of the interdigital diamond detector is high for particles with energies low enough to be completely stopped inside this high-efficiency layer, and it starts to decrease for those having higher energies, and thus larger ranges, up to the case where they will generate a signal on the detector comparable to the noise level. Indeed, it is well known that the charge collected in diamond detectors due to $N_i$ incident particles, depositing energy $E_i$ in the sensitive layer, can be expressed by $Q_i = \eta_i(E_i) q_e N_i E_i / \epsilon$, where $q_e$ is the electron charge, $\epsilon$ is the average electron-hole pair energy creation, i.e. 13 eV for the diamond, and $\eta_i \leq 1$ the Charge Collection Efficiency (CCE) of the detector for a given particle[42].

In order to evaluate the CCE curve, the diamond detector was previously characterized at the AN2000 microbeam facility of the National Laboratories of Legnaro (Italy) by using proton beams of energy respectively 0.3 MeV, 1 MeV, 1.4 MeV and 2 MeV. The beam was focused to a spot size of ~5 µm on the sensitive electrodes of diamond. As evaluated by the SRIM Monte Carlo Simulation code[26], the range of protons in diamond varies from 1.5 µm to 25 µm. An electronic chain consisting of a charge-sensitive preamplifier ORTEC142A and an ORTEC572 shaping amplifier was used. The calibration of the electronic chain was performed using a Si detector and a precision pulse generator ORTEC419 relating the pulse heights provided by the reference Si detector with those from the diamond devices.

The CCE of the detector as a function of the investigated proton energies is reported in Figure 5.

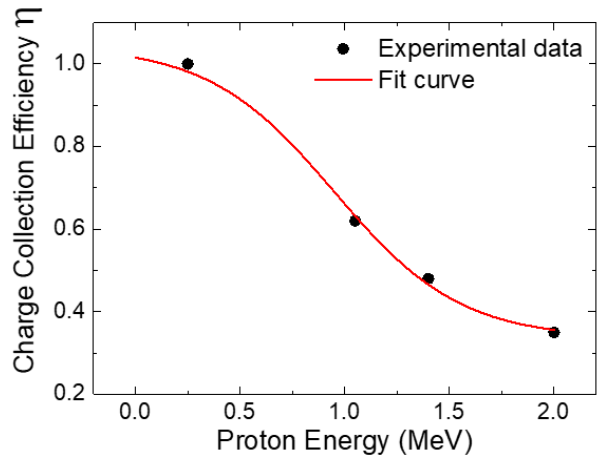

**Figure 5** Charge Collection Efficiency (CCE) of the diamond detector as a function of energy. Fit curve is also reported in the figure.

As expected, for low energy protons, the induced charge is equal to the generation charge (i.e. $\eta = 1$) whereas, for high energy protons, charges are generated in the region where weak electric field is present, and therefore the CCE of the detector decreases. The probability for free carriers to enter the depletion region exponentially decreases, with a logarithmic slope equal to their diffusion length. The experimental data are well fitted by a non-linear curve having the following expression:

$$\eta_i(E_i) = \frac{0.71}{1 + e^{\frac{E_i - 9.1 \cdot 10^5}{3.1 \cdot 10^5}}} + 0.33 \tag{5}$$

By considering that equal CCE is obtained for other particles having the same range in the detector[41], from these measurements it is possible to determine the CCE also for other ion species.
The number of incoming ions of known specie can be retrieved according to the following relation:

$$N_i = \frac{Q_i \, \epsilon}{\varphi(E_i) \, q_e} \tag{6}$$

where the definition $\varphi(E_i) \equiv \eta_i(E_i) \, E_i$ is used for simplicity. In a generic $n^{th}$ time interval, the value of the associated charge generated in diamond can be computed:

$$\bar{Q}_{i,n} = \frac{1}{R} \int_{t_{i,n}^i}^{t_{i,n}^f} S_D[V(t)] dt \tag{7}$$

where $R$ is the characteristic load impedance and $S_D[V(t)]$ is the de-embedded signal. Also for the number of incoming ions it is possible to determine the uncertainty in the determination of $N_i$. In particular, equation (6) can be modified as follows:

$$\bar{N}_{i,n} = \frac{\bar{Q}_{i,n} \, \epsilon}{\bar{\varphi}_{i,n} \, q_e} \tag{8}$$

where $\bar{N}_{i,n}$ and $\bar{\varphi}_{i,n}$ are the average of the number of ions intercepting the detector in the $n^{th}$ time interval, and the related average of the $\varphi$ function in the same interval. As a consequence, the relative error on the number of incoming ions, given by the finite temporal resolution, can be written as:

$$\frac{\Delta N_{i,n}}{N_{i,n}} = \frac{N_{i,n}(E_{i,n}^f) - N_{i,n}(E_{i,n}^i)}{\bar{N}_{i,n}} = \bar{\varphi}_{i,n} \left[ \frac{1}{\varphi(E_{i,n}^f)} - \frac{1}{\varphi(E_{i,n}^i)} \right] \tag{9}$$

Hence, the error on the estimation of the number of impinging particles can be expressed in terms of the $\varphi$ function and is thus strongly related to the temporal resolution of the detector. Nevertheless, it is important to underline that the performed evaluation holds on the assumption that only one kind of particle is entering the detector.

**Time superimposition of detected particles**
In laser-matter interaction experiments, different ion species are classically accelerated and arrive to the detector usually superimposed in time. Indeed, in a classical TNSA scheme, ions are accelerated because of the electrostatic potential developed in the charged sheath formed in the target thanks to the electron emission[2-4,49]. Because of this potential, ions are accelerated with a typical spectrum represented by a decreasing exponential function[4]. A broad energy distribution is given by the acceleration of particles from different target depths and by the inhomogeneous electron distribution in the sheath[4,49]. As a general discussion, for a test ion accelerated by the $\Phi$ potential drop in the sheath, having $A_i$ mass number, $z_i$ charge state and $Z_i$ atomic number, it is possible to write that the energy is $E_i = z_i\, q_e\, \Phi$, where $q_e$ is the elementary charge[3]. For ion energies up to several tens of MeVs non-relativistic formulas can be used, and thus the ion time-of-flight at the diamond detector will be:

$$t_i = d\sqrt{\frac{A_i m_p}{2 z_i q_e \Phi}} = t_p \sqrt{\frac{A_i}{z_i}} \qquad (10)$$

where $m_p$ is the proton mass and $t_p$ the time of arrival to the same detector of a proton accelerated by the same potential drop. Since for any ion, except for protons, it is $A_i > z_i$, then it will always be $t_i > t_p$. Thus, protons will reach the detector much before than other ions accelerated in the same experiment. Moreover, there will be a time window where only protons can reach the detector. If we indicate with $t_{pM}$ and $t_{iM}$ the times of arrival to the detector of the proton with the maximum energy and of the corresponding ion with the maximum energy, generated respectively by the same acceleration process, the time window will be $(t_{pM}, t_{iM}) = (t_{pM}, t_{pM} - \Delta t_p)$. In the limiting case of fully-stripped ions (i.e. $z_i = Z_i$), it is possible to write:

$$\Delta t_p = t_{pM}\left(1 - \sqrt{\frac{A_i}{Z_i}}\right) \qquad (11)$$

It is well known that for any element (with exception of protons), it is $A_i \geq 2Z_i$, being the equality true for light elements (C, N, O, …) classically accelerated in intense laser-matter experiments. So, it is:

$$\Delta t_p \geq t_{pM}(1 - \sqrt{2}) \qquad (12)$$

It is possible to determine the same condition in the energy domain. So, the proton energy range associated with the extremes of the time interval of equation (11) can be written as $(E_{iM}, E_{pM}) = (E_{pM} - \Delta E_p, E_{pM})$, where:

$$\Delta E_p = E_{pM}\left(1 - \frac{Z_i}{A_i}\right), \qquad (13)$$

And for $A_i \geq 2Z_i$:

$$\Delta E_p \geq \frac{E_{pM}}{2}, \qquad (14)$$

So, for a given experiment of TNSA mechanism, or in general for any laser-matter process where ions are accelerated because of a potential drop, we can consider as a general criterion that if in a TOF scheme we are able to detect the maximum proton energy $E_{pM}$, then in at least the detected proton energy range $(E_{pM}/2; E_{pM})$ there will be no contribution coming from the superimposition of other ions. It is worth to consider that this estimation was achieved under the rather conservative hypothesis of fully stripped light ions, which is a very limiting case, since in typical experiments $z_i < Z_i$ for the largest part of the ions.

**Filter application**

In some shots of the present campaign an Aluminium filter of $(10 \pm 1.5)$ μm thickness was placed in front of the interdigital diamond detector. Filters are usually employed to cut the contribution of heavier ions from the detected signal, with a little loss of information on protons. We define $E_{i-cutoff}$ the maximum energy of ions of a given species fully stopped by this specific filter. According to the simulations performed with the SRIM code, for the nominal 10 μm thickness of the used filter, it was found $E_{p-cutoff}$ = 750 keV for protons[50] (see Figure 6a). Simulations also gave cutoff energies of 11.45 MeV for Carbon, 13.7 MeV for Nitrogen and 15.84 MeV for Oxygen ions, the main species expected to be accelerated in this type of experiments as impurities on the target surface, and 25.2 MeV for Aluminium ions coming from the bulk target. Moreover, in Figure 6a the energy attenuation for protons able to pass through the filter was computed for the same 10 μm thickness and for the minimum and maximum values of thickness according to the filter tolerance (8.5 μm and 11.5 μm respectively). This attenuation factor is obtained by the relation $k_{att}(E_{in}) = E_{out}/E_{in}$ where $E_{in}$ is the energy of the particle impinging on the filter and $E_{out}$ is the energy of the particle that crossed the filter.

In general, to retrieve the proton spectrum in case a filter is used between the target and the detector, it is necessary to correct the number of particles estimated from equation (6) as it follows:

$$N_i = \frac{Q_i \epsilon}{k_{att}(E_i)\, \varphi(E_i)\, q_e} \quad (15)$$

and also the related equations (7) and (8) accordingly, where the attenuation factor $k_{att}(E_i)$ was here retrieved by SRIM simulations. Moreover, in a classical TOF scheme, the particle energy is easily linked to a specific time instant from the equation of uniform rectilinear motion. When a filter is used, the overall path is schematically split in two consecutive ones, where the same particle has different velocities. The first, having length D, from the target to the filter, and the second of length δ, from the filter to the detector. In general, the overall arrival time $\bar{t}_{p,n}$ of a proton to the detector can be thus written as:

$$\bar{t}_{p,n} = \bar{t}_{pD,n} + \bar{t}_{p\delta,n} = D\sqrt{\frac{m_p}{2\bar{E}_{p,n}}} + \delta\sqrt{\frac{m_p}{2k_{att}(\bar{E}_{p,n})\bar{E}_{p,n}}} = \sqrt{\frac{m_p}{2\bar{E}_{p,n}}}\left(D + \frac{\delta}{\sqrt{k_{att}(\bar{E}_{p,n})}}\right) \quad (16)$$

This relation actually links arrival time and energy in the case a filter is used, and can be inverted numerically to find the associate $\bar{E}_{p,n}$, since $k_{att}$ is a monotonic function to be obtained by offline experimental calibration of the filter or numerical simulations. It is clear that the second term will be of interest only for particles with energy close to the filter cutoff. As a rule of thumb we can consider that the effect of the drift space located after the filter starts to be relevant when the time needed by a particle to travel through it is one tenth of the time needed to travel the main drift space D. To evaluate for which proton energy this happens it is convenient to express the filter attenuation factor as a function of the parameter $G = \bar{t}_{p\delta,n}/\bar{t}_{pD,n}$:

$$k_{att}(\bar{E}_{p,n}) = \left(\frac{\delta}{DG}\right)^2 \quad (17)$$

In the present experiment $\delta = 0.2$ cm and $D = 105$ cm; a value of $G = 0.1$ was thus achieved for $k_{att} = 3.64 \times 10^{-4}$. This value of the attenuation factor, as shown in Figure 6a, is obtained for those protons with energies near the filter cutoff or lower than that.

We can conclude that in the present study the effects of the drift $\delta$ can be neglected, but it is important to take into account the energy attenuation in equation (15). Nevertheless, equation (16) is very useful for all those cases where the filter is placed not too close to the detector, or when it is important to have rather accurate ion energy estimations very close to the filter cutoff.

**RESULTS**

The described procedure was applied to retrieve the particle spectrum under the preliminary assumption that all incoming particles were protons. The proton energy spectrum reconstructed following the proposed procedure from the TOF signal shown in Figure 2b, i.e. shot #38 ($E_L = 2.5$ J, $I_L = 1.7 \times 10^{19}$ Wcm$^{-2}$) is reported in Figure 6b. From this spectrum it is also possible to achieve, under the assumption of uniform angular emission, rough estimations of the total number of generated protons $N_{p,tot}$, the associated total energy $E_{p,tot}$ and the laser-to-proton energy conversion factor $C_{L\to p}$. The following parameters were extracted: $N_{p,tot} = 2 \times 10^{13}$ sr$^{-1}$, $E_{p,tot} = 0.5$ J and $C_{L\to p} = 20\%$. These values are far above those known from the literature[51,52].

This overestimation can be easily explained by taking into account that other particles besides protons contributed to the detected signal, as discussed in the previous section. In particular, by applying those considerations to the same shot #38, under the conservative assumption of fully stripped light ions reaching the diamond, it results ($t_{pM} - \Delta t_p$, $t_{pM}$) = (65.4 ns, 92.4 ns) or, in terms of proton energy, ($E_{pM} - \Delta E_p$, $E_{pM}$) = (1.077 MeV, 2.154 MeV). By limiting the estimation of $N_{p,tot}$, $E_{p,tot}$ and $C_{L\to p}$ to this interval, the values $9 \times 10^{10}$ sr$^{-1}$, 0.02 J and 0.8 % are obtained, respectively. These are sensibly lower than those obtained in the previous evaluation and are in good agreement with those reported in literature for similar experimental conditions[51,52].

The analysis of the data collected from those shots where an aluminium filter was used provided a further confirmation of the obtained results and of the developed theoretical evaluations. In particular, the de-embedded signal obtained for the shot #5 ($E_L = 2.65$ J, $I_L = 1.8 \times 10^{19}$ Wcm$^{-2}$), having similar conditions to the previous #38, is reported in Figure 7a. In the white region on the left, particles reaching the detector do not feel the effect of the filter remarkably, which is indeed effectively acting on those arriving at later times, i.e. in the red region. In particular, the signal detected for t > 100 ns is given by two main contributions. The first is from the bremsstrahlung radiation from the particles interacting with the filter. The second is the contribution of the particles with energies close to the cutoff, which are still able to pass through the filter, which strongly reduces their velocity and also their number, as described by equation (16). The spectrum obtained for protons in the white region of Figure 7a is illustrated in Figure 7b. For shot #5 we obtain, $N_{p,tot} = 1.87 \times 10^{11}$ sr$^{-1}$, $E_{p,tot} = 0.03$ J and $C_{L\to p} = 1\%$, in very good agreement with the previous case without the filter, when the simultaneous particle detection is properly considered.

Nevertheless, the obtained spectrum is not completely free from heavier ion contribution. The simultaneous detection of protons and ions is the reason for the evident rise at low energies of the computed spectra in Figure 7. This is not related to the proton acceleration process itself but is due to the contribution to the spectrum of Carbon ions precisely in that energy range, which cannot be separated from the one of the protons. According to the previous discussion, for all those energies lower than $E_{p,M}/2$ the signal could be given by the sum of ions and protons impinging on the diamond detector. For instance, using the relation $E_C = m_C/m_p \cdot E_{p,M}/2$ it is possible to compute the equivalent energy that a carbon ion detected with the same delay as a certain proton would have. For the spectrum given in Figure 7b we have $E_{p,M}/2 = 1.153$ MeV, which means that if the signal were generated by Carbon ions, they would have an energy $E_C = 13.8$ MeV. As already mentioned the cutoff energy of the 10 µm Aluminium filter for Carbon ions is 11.5 MeV, hence we have a window where Carbon ions are able to pass through the filter and reach the diamond detector, namely from 11.5 MeV to 13.8 MeV or, in terms of protons' energy, from 0.953 MeV to 1.153 MeV. Actually, as already discussed for protons, the Carbon ion passing through the filter will be slowed down, resulting in an additional delay between different ion species. For instance, in the case under investigation we see that, using equation (16), the time needed for protons of 1.153 MeV to reach the detector is $\simeq 9.1$ ns, while the time needed for Carbon ions of equivalent energy is $\simeq 11.3$ ns, which means that they are going to affect the signal generated by protons of 0.906 MeV. In other terms the spectral range affected by Carbon ions results to be shifted from (0.953, 1.153) MeV to (0.750, 0.906) MeV. Similar consideration should be applied for heavier ions.

Figures 7b-d show the proton spectra obtained for the shot #5 using the nominal thickness of the filter and its upper and lower limit on the basis of the tolerances, i.e. 8.5 µm and 11.5 µm. The increase of the error bar on the amplitude is due to the associated tolerance in the attenuation of the aluminium filter given in Figure 6a and it is more relevant for low energy particles. Low-energy particles are those that are more affected by the presence of the filter and, especially near to the filter cut-off energy, a small difference in its thickness can result into a quite big difference in the respective attenuation factor.

In Figure 8, all the mentioned spectra for shot #38 and for shot #5 are compared on the same graph, without error bars for a better visualization. It is possible to notice that the shots were rather similar in terms of spectrum, and the use of the filter allowed to down-extend the proton energy range, at expenses of the larger errors introduced in the lower energy part of the obtained spectrum.

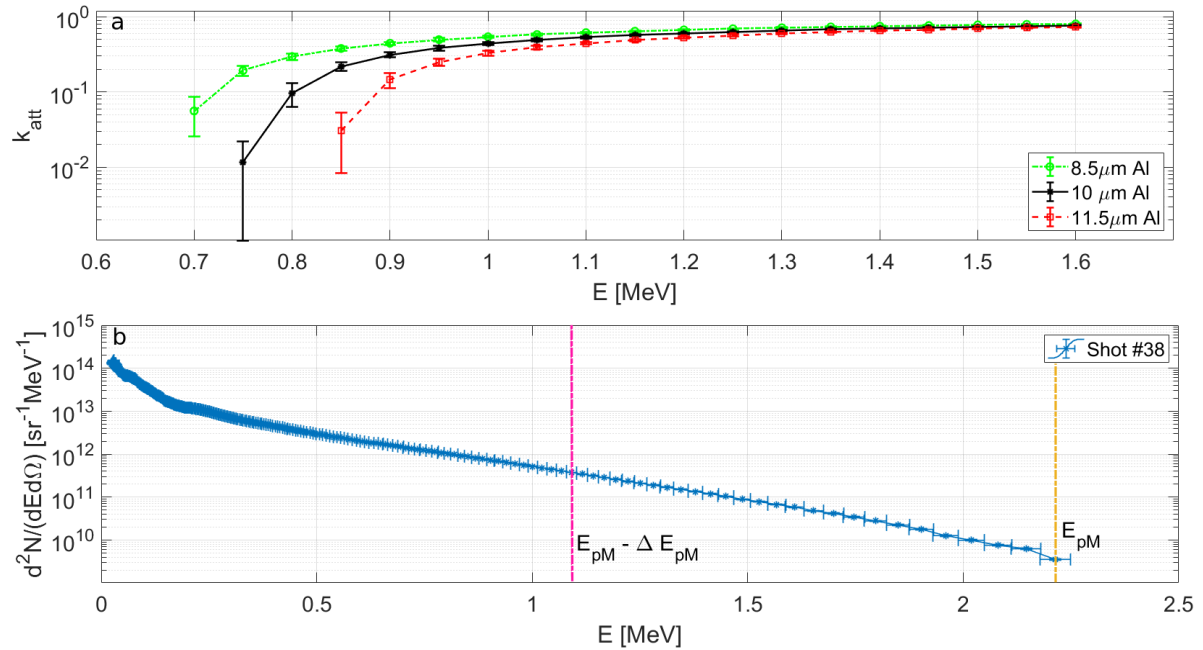

**Figure 6.** (a) Attenuation factor $k_{att}$ for the (10 ± 1.5) µm Aluminium filter obtained from SRIM simulations; (b) proton spectrum obtained for the Shot #38 with the described

procedure. The vertical yellow line indicates the maximum proton energy ($E_{pM}$) and the magenta line corresponds to the energy where the ion contribution to the spectrum ends ($E_{pM} - \Delta E_{pM}$)

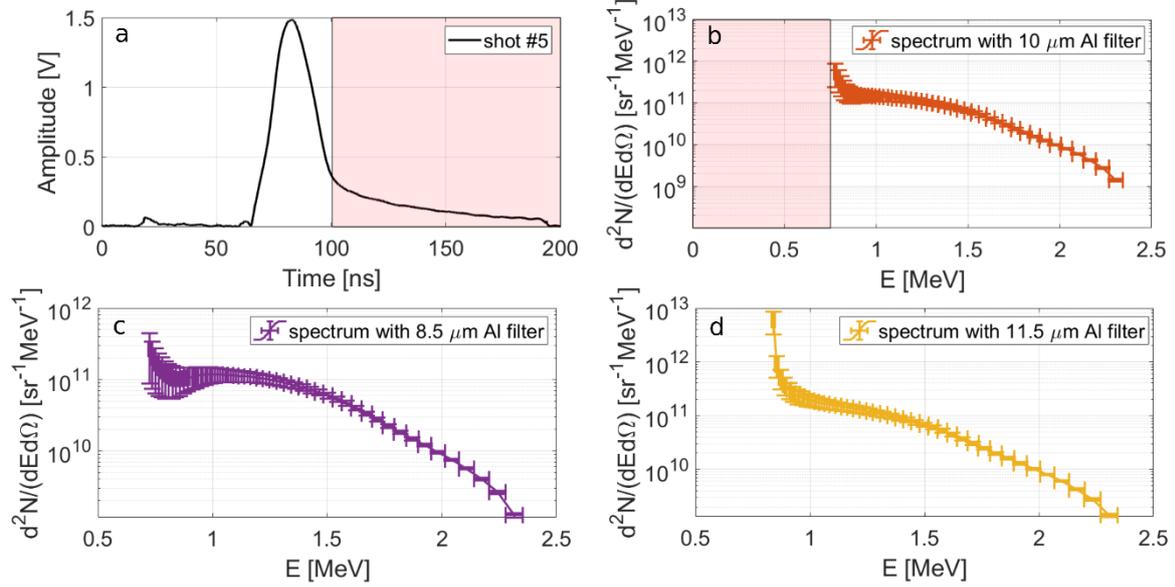

**Figure 7.** (**a**) Signal obtained for shot #5 with the Al filter covering the diamond detector. (**b-d**) Proton spectrum obtained for this shot by taking into account the effect of the Al filter of thickness 10 µm (b), 8.5 µm (c) and 11.5 µm (d).

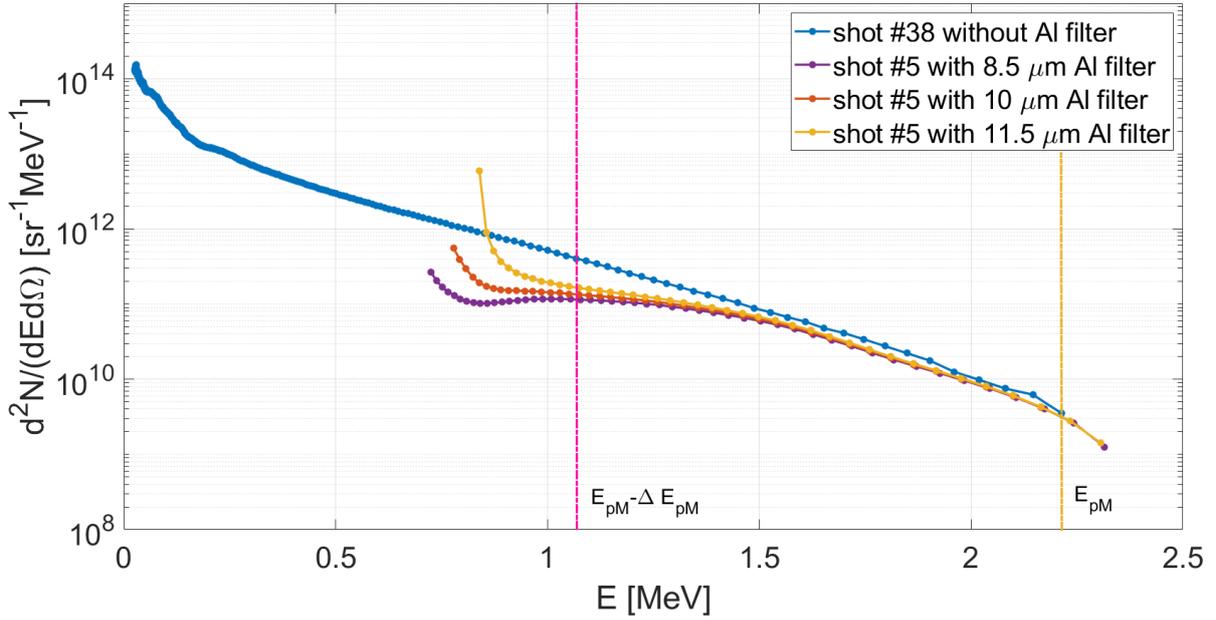

**Figure 8.** Comparison between the spectra obtained from the signals acquired for the shot #38, without the Al filter, and shot #5 with the Al filter for the three thicknesses, according to the nominal tolerances. The vertical yellow line indicates the maximum proton energy ($E_{pM}$) obtained during shot #38 and the magenta line corresponds to the energy where the ion contribution to that spectrum ends (namely $E_{pM} - \Delta E_{pM}$)

**DISCUSSION**

In this work we reported on the details of an effective and accurate TOF measurements in a highly EMP-polluted environment. The described methodology allowed to optimize the experimental set up and the acquisition system in order to get a high signal-to-noise ratio on the whole dynamic range of the collected signals; this was achieved thanks to the high rejection to the EMP fields reached in the described way. The tailored optimization of the analogic signal management enabled the retrieving of both the interaction-time and the particle information from the recorded data with high accuracy.

The TOF method was here used to get information on the pure proton contribution from all the particle species detected by diamond detector. Measurements and data retrieving were performed promptly for every shot, without any need of opening the vacuum chamber. Detailed considerations were here presented for getting the calibrated proton spectrum associated to the time-domain measurements. The results of the theoretical discussion were applied to experimental data for shots performed with ~120 TW power and $2 \times 10^{19}$ Wcm$^{-2}$ intensity. A novel detailed procedure was applied for getting accurate spectra from the diamond detector, by taking into account its efficiency as a function of the energy of the impinging particle, obtained by the experimental calibrations of its response to protons of different energy.

Moreover, it was here shown that the use of tailored filters can positively decrease the lower value of proton spectrum, in general at some expenses of the tolerance. This can be anyway improved for filters where the thickness is known or determined with high accuracy. In the described experimental campaign, the advanced TOF method allowed to obtain calibrated spectra of protons up to ~2.5 MeV with very high accuracy, much better than that typically obtained with Thomson spectrometers[25,53].

The capability of performing such on-line accurate measurements in these environments with high EMP levels but without the use of Faraday cages, is of crucial importance to monitor and characterize effectively the particles accelerated via laser-plasma interaction in high-repetition laser-matter experiments. This is a key point for high-intensity and high-energy laser facilities for both laser-plasma acceleration and inertial confinement fusion (PETAL, Vulcan Petawatt, Extreme Light Infrastructure (ELI)), where high levels of EMP are classically produced. Since the EMP fields are known to scale with laser energy and intensity[33], the problem will be of increasing importance for future facilities which will be operating soon (Apollon 5 PW, ELI L4 10 PW, …). where higher EMP levels are expected. The technique was tested at FLAME laser and, in order to understand its possible application to facilities affected by even higher EMP pollution levels, some preliminary and successful tests were performed in experiments at the Vulcan Petawatt laser[54] with laser energy $E_L = 500$ J, pulse width 500 fs, $I_L \cong 10^{21}$ Wcm$^{-2}$, and also very recently at the Phelix laser with $E_L = 100$ J, pulse width 750 fs, $I_L \cong 10^{20}$ Wcm$^{-2}$ (these latter results will be presented in a separate publication).

The potential high energy resolution is one of the key factors of the TOF technique and was here proved by using thin-fast diamonds in an advanced implementation of this scheme. By using these advanced and optimized methodologies, fast diamond detectors can be thus successfully adopted as a main constituent of a layered diamond detector system made of a stack of them. Such configuration will allow to get a composite detector with effective thickness of extended dimension without losing energetic resolution and featured by high radiation hardness, thanks to the independent thin diamonds available throughout the structure. These considerations confirm that the described methodology together with the overall advanced associated detector system is a promising candidate for the fast, accurate, high-resolution, high sensitivity, high radiation hardness and online detection of energetic ions in experiments of high energy and high intensity matter interaction with ultra-short laser pulses up to the femtosecond and even future attosecond range.

**Acknowledgements**
The work has been partially carried out within the framework of the EUROfusion Consortium and has received funding from the Euratom research and training program 2014–2018 and 2019–2020 under grant agreement No. 633053. The views and opinions expressed herein do not necessarily reflect those of the European Commission.

**Author Contributions**
F.C. conceived the idea of the advanced TOF methodology and discussed the details with C.V., M.S. and M.C.; M.S., F.C., M.C. and C.V. performed its development and the related theoretical and numerical considerations, with contributions by R.D.A., P.L.A., G.Cr., G.D.G. and valuable advices by M.M., P.A. and D.G. . C.V. developed and realized the diamond detector, performed the calibrations and determined the efficiency. The TOF measurements were performed by M.S., F.C., M.C., together with M.P.A., F.B., G.Co. and A.Z.
M.S. analyzed the TOF data, performed the associated computations, prepared the figures and wrote the manuscript under the supervision of F.C., with valuable advices by M.C., C.V., M.M., P.A., R.D.A., D.G. .
A.Z. and M.F. planned and managed the laser-foil experiment. M.P.A., F.B., G.Co. and M.G. managed the FLAME laser and the related diagnostics during the experiment. M.P.A., F.B., G.Co., R.P., M.G. and A.Z. carried out the experimental campaign studying laser interaction with foil targets that provided the experimental platform for the work. A.Z. was the lead researcher on the experimental campaign.
All authors discussed the results and reviewed the manuscript.

**Additional Information**
**Competing interests**
The authors declare no competing interests

**Correspondence** and requests for materials should be addressed to M.S.